\documentclass[twoside,onecolumn]{article}

\usepackage{blindtext} 

\usepackage[sc]{mathpazo} 
\usepackage[T1]{fontenc} 
\usepackage{microtype} 
\usepackage[english]{babel} 
\usepackage[hmarginratio=1:1,top=32mm,columnsep=20pt]{geometry} 
\usepackage[hang, small,labelfont=bf,up,textfont=it,up]{caption} 
\usepackage{booktabs} 
\usepackage{lettrine} 
\usepackage{enumitem} 
\setlist[itemize]{noitemsep} 
\usepackage{abstract} 
\usepackage{titlesec} 
\renewcommand\thesection{\Roman{section}} 
\renewcommand\thesubsection{\roman{subsection}} 
\titleformat{\section}[block]{\large\scshape\centering}{\thesection.}{1em}{} 
\titleformat{\subsection}[block]{\large}{\thesubsection.}{1em}{} 
\usepackage{fancyhdr} 
\usepackage{graphics}
\usepackage{graphicx}
\usepackage{epsfig}
\usepackage{amssymb}
\usepackage{cite}
\usepackage{titling} 
\usepackage{hyperref} 
\usepackage{ragged2e}
\usepackage{amsmath}
\usepackage{mathptmx}
\usepackage{graphicx}  
\usepackage{dcolumn}   
\usepackage{bm}        
\usepackage{amssymb}   
\usepackage{subfigure}  
\usepackage{hyperref}
\usepackage{ragged2e}
\usepackage{mathtools}
\usepackage{xcolor}
\usepackage{nccmath}
\usepackage{ragged2e}
\usepackage{graphics}
\usepackage{graphicx}
\usepackage{epsfig}
\usepackage{amssymb}
\usepackage{lineno}
\usepackage{blindtext}
\usepackage{lineno}
\usepackage{lineno, blindtext}
\usepackage{mathtools}
\usepackage{xcolor}
\usepackage{hyperref}
\usepackage{microtype}
\usepackage{breakcites}
\usepackage{footnote}
\usepackage{appendix}
\usepackage{ragged2e}

\pretitle{\begin{center}\LARGE\bfseries} 
\posttitle{\end{center}} 
\title{High yield, shell-stabilized, narrow-sized $C_3F_8$ nanobubbles with different shell properties and precisely controllable response to acoustic excitations: experimental observations and numerical simulations} 
\author{%
\textsc{AJ. Sojahrood $^{a,b}$ \thanks{AJ. Sojahrood and Al de Leon contributted equally and sahre the first authorship. Email: amin.jafarisojahrood@ryerson.ca}} ,\textsc{Al de Leon$^c$}, \textsc{Richard Lee$^d$, Michaela Cooley$^c$}  \\[1ex] %
\textsc{ Eric Abenojar$^c$, Michael C. Kolios $^{a,b}$ \thanks{Corresponding author, Email: mkolios@ryerson.ca} and Agata A. Exner $^{c}$ \thanks{Corresponding author, Email: agata.exner@case.edu}}\\
\normalsize $^a$ Department of Physics, Ryerson University, Toronto, Ontario, Canada. \\
\normalsize $^b$ Institute for Biomedical Engineering and Science Technology, \\
\normalsize A Partnership Between Ryerson University and St. Michael’s Hospital, Toronto, Canada. \\
\normalsize $^c$ Department of Radiology Case Western Reserve University, Cleveland, OH, 44106, USA\\
\normalsize $^d$Light Microscopy Imaging Core, Case Western Reserve University, Cleveland, OH, 44106, USA\\ }
\date{\today} 
\renewcommand{\maketitlehookd}{
\begin{abstract}
Understanding the pressure dependence of the nonlinear behavior of ultrasonically excited phospholipid (PL)-stabilized NBs is important for optimizing US exposure parameters for implementations of contrast enhanced ultrasound, critical to molecular imaging. The viscoelastic properties of the shell can be controlled by introduction of membrane additives, such as propylene glycol as a membrane softener or glycerol as a membrane stiffener. We report for the first time, the production of high yield NBs with narrow dispersity and different shell properties. Through precise control over size and shell structure, we show how these shell components interact with the phospholipid membrane, change its structure, affect their viscoelastic properties, and consequently change their acoustic response. A two-photon microscopy technique through a polarity-sensitive fluorescent dye, C-laurdan, was utilized to gain insights on the effect of membrane additives to the membrane structure. We report how the shell stiffness of NBs affects the pressure threshold ($P_t$) for the sudden amplification in the scattered acoustic signal from NBs. For narrow size NBs with 200 nm mean size, we find $P_t$ to be between 123-245 kPa for the NBs with the most flexible membrane as assessed using C-Laurdan, 465-588 kPa for the NBs with intermediate stiffness and 588-710 kPa for the NBs with stiff membranes. Numerical simulations of the NB dynamics are in good agreement with the experimental observations confirming the dependence of acoustic response to shell properties, thereby substantiating further the development in engineering the shell of UCAs.  The viscoelastic dependent threshold behavior can be utilized for significantly and selectively enhancing the diagnostic and therapeutic ultrasound applications of potent narrow size NBs. 
\end{abstract}
}

\begin{document}

\maketitle


\section{Introduction}
\justify
Clinical ultrasound contrast agents (UCAs), also referred to as microbubbles (MBs), have augmented the capabilities of ultrasound (US) in areas such as cancer detection, tumor characterization, and theranostics \cite{1,2,3,4,5}.  There has been a substantial recent interest in the pre-clinical development of novel, nanoparticle-based UCAs; these include nanobubbles (NBs), nanodroplets, and nanovesicles \cite{6,7,8,9,10}. One advantage of the sub-micron UCAs is that they have been shown to extravasate beyond leaky tumor vasculatures, unlike MBs that are confined to the blood vessels because of their large size (1-$10\mu m$) \cite{11,12,13,14,A}. This extravasation is well-suited for applications such as molecular imaging and targeted drug delivery. Applications of sub-micron UCAs range from measuring T lymphocyte infiltration in cardiac tissue \cite{15} to detection of type 1 diabetes \cite{B}, prostate cancer \cite{13,16} and targeted delivery for photothermal therapy \cite{17}. Despite the recent growth of MB and nanobubbuble-based imaging applications, little work has been done thus far in understanding how the physical properties of the shell determine their interaction with US, and whether this interaction is consistent with the current theoretical and experimental understanding of models of bubble oscillation. In this work we thus examine how changes in the NB size distribution and shell structure affect their acoustic response.\\
The dynamics of MBs in an acoustic field depend strongly on US parameters (e.g. US acoustic pressure and frequency) and bubble properties (e.g. size, gas, shell elasticity, and shell viscosity), and can be mathematically described by nonlinear encapsulated bubble models such as the Marmottant model \cite{18,19,20}. Numerous studies have demonstrated the strong effect of the UCA shell elasticity and size on the UCA resonance frequency \cite{18,21,22}, and acoustic pressure of maximal signal intensity with minimal microbubble destruction \cite{23}. Experiments with various lipid shell compositions have also shown a strong dependence on nonlinear MB behavior \cite{24,25,26,27,28}. Thus, the rational design of the shell structure and size of the UCAs has the potential to tune their behavior to a given ultrasound frequency and pressure. \\
The shell properties of phospholipid (PL)-stabilized UCAs can be altered by introducing membrane additives. PL shells can be made stiffer by incorporating membrane stiffeners such as glycerol (Gly) and carbohydrates, or more flexible by incorporating membrane softeners (or edge-activator) such as propylene glycol (PG) and cholesterol \cite{29,30,31,32}. Gly has been shown through X-ray and neutron reflectivity measurements to preferentially interact via hydrogen bonding with the PL head, dehydrating the PL shell, and increasing shell stiffness \cite{33}. On the other hand, PG has been utilized as a membrane softening component in ultradeformable liposomes \cite{34,35}. PG assembles in the PL membrane, reducing PL packing order and stiffness, and imparting membrane fluidity \cite{34,35,36,37,38,39,40}. Incorporation of either Gly or PG into a bubble shell affects its shell properties as a result of a change in PL packing order \cite{33,34,39,41,42,43}. \\
Despite the importance of shell properties on the behavior of MBs and NBs, however, to our best knowledge no experimental studies have explored the effect of shell structure on the bubble behavior - independent of the effect of the bubble size distribution. In polydisperse solutions, the acoustic response is dominated by bubble-to-bubble variations that would dominate over any effects of shell structure. In this work, we aim to investigate the effect of shell stiffness on the nonlinear behavior of NBs independent of the size effects.  To achieve this, first NBs with 3 different shell compositions were manufactured.  NBs of different shell stiffness were prepared by incorporation of different amounts of Gly as a membrane stiffener and PG as a membrane softener. The relative PL packing order in the bubble membrane was assessed by a common assay typically used to examine lipid packing in cell membranes \cite{30,44,45,46}. The technique provides complementary information to the developed shell property measurement techniques.   This can be done through two-photon microscopy with a polarity-sensitive fluorescent probe such as C-laurdan  (6-lauryl-2-dimethylamino-napthalene) \cite{30,31,44,45,46,47,48,49,50} by calculating the average GP value from the emitted fluorescence intensities at 450-nm and 500-nm after exciting C-laurdan with a 800-nm laser in a two-photon microscopy set-up. This method has been used to measure lipid transfer from MBs to cell membranes, and recently to measure MB shell characteristics \cite{51}. \\
We then introduce a simple but effective method to produce NBs with very narrow size distribution and high yield. Three NB populations were filtered to have similar sizes with a narrow size distribution and diluted to have similar concentration.  The effect of shell characteristics on the nonlinear oscillations of NBs in an US field was then studied by exposing NB solutions to US of varying pressures and analyzing the contrast harmonic images. The dependence of pressure for substantial increase in nonlinear oscillation of PL-stabilized NB solution (~200 nm diameter) on shell stiffness was studied both experimentally and numerically.
\section{Materials and methods}
\subsection{Materials}
\justify

1,2-dibehenoyl-sn-glycero-3-phosphocholine (C22, Avanti Polar Lipids Inc., Pelham, AL), 1,2 Dipalmitoyl-sn-Glycero-3-Phosphate (DPPA, Corden Pharma, Switzerland), 1,2-dipalmitoyl-sn-glycero-3-phosphoethanolamine (DPPE, Corden Pharma, Switzerland), and 1,2-distearoyl-snglycero-3-phosphoethanolamine-N-[methoxy(polyethyleneglycol)-2000](ammonium salt)
(DSPE-mPEG2000, Laysan Lipids, Arab, AL), propylene glycol (PG), glycerol (Gly), phosphate buffer solution (PBS, Gibco, pH 7.4), 6-Dodecanoyl-N,N-dimethyl-2-naphthylamine (C-Laurdan, Sigma Aldrich), octafluoropropane ($C_3F_8$, Electronic Fluorocarbons, LLC, PA), Agarose (Sigma Aldrich).
\subsection{Preparation of bubble solutions}
\justify
Nanobubbles were formulated as reported previously \cite{52,53}.  Briefly, solution for bubbles with membrane of intermediate flexibility (10 mg/mL) was prepared by first dissolving 6.1 mg C22, 1 mg DPPA, 2 mg DPPE, and 1 mg DSPE-mPEG2000 into 0.05 mL PG by heating and sonicating at 80$^{0}$C until all the lipids were dissolved. Mixture of 0.05 mL Gly and 0.9 mL PBS preheated to 80$^{0}$C was added to the lipid solution. The resulting solution was sonicated (Branson Sonicator CPX2800H) for 10 min at room temperature. The solution (1 mL) was transferred to a 3 mL headspace vial, capped with a rubber septum and aluminum seal, and sealed with a vial crimper. The solutions for bubbles with flexible and stiff membranes were prepared similarly but with 0.1 mL of PG or 0.1 mL of Gly, respectively, added to the solution instead of 0.05 mL of PG and 0.05 mL of Gly.
\begin{figure}
	\begin{center}	
		\includegraphics{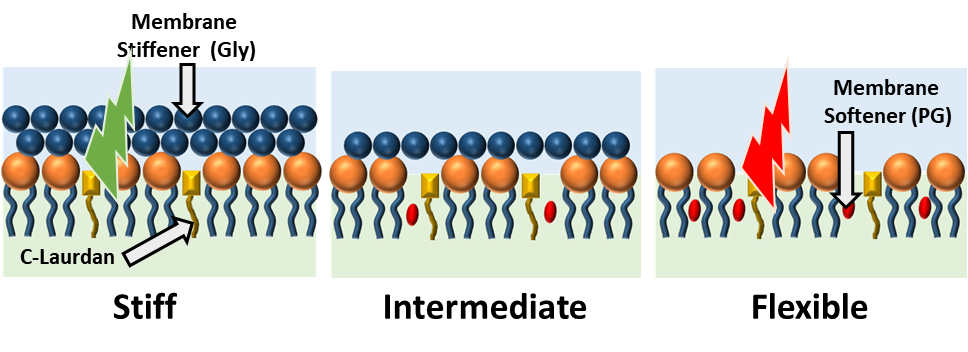}
		\caption{Schematic of bubble membrane showing the influence of membrane stiffener and membrane softener in the PL packing as detected by the fluorescence emission of C-laurdan. }
	\end{center}
\end{figure}
\subsection{Quantitative imaging of membrane lipid order with C-Laurdan}
The relative change in PL packing order and stiffness upon incorporation of additives was determined through quantitative two-photon fluorescence microscopy with a polarity-sensitive fluorescent probe (C-laurdan). 5 $\mu$L of 5 nM C-laurdan solution in DMSO was added to each bubble solution. To form microbubbles, air was manually removed with a 30 mL syringe and was replaced by injecting $C_3F_8$ gas. After air was replaced by $C_3F_8$, the phospholipid solution was activated by mechanical shaking with a VialMix shaker (Bristol-Myers Squibb Medical Imaging Inc., N. Billerica, MA) for 45s. 0.1 mL of bubble solution was withdrawn and mixed with 1 wt$\%$ agarose solution in PBS at 30 $^{0}$C. 100 $\mu$L of agarose solution with bubbles was transferred to a glass bottom dish for two-photon microscopy imaging using a Leica TCS SP2 multiphoton confocal system (Buffalo Grove, IL) equipped with a Coherent Chameleon XR IR laser (Santa Clara, CA) tuned to 800 nm. Samples were imaged using either a 63x/1.40 NA Oil or a 63x/1.20 NA water immersion objective. Sixty bubbles for each bubble type were imaged with a sampled pixel size of ca. 230 nm, using 2-line averages and 2 frame averages. The two-photon microscope was pre-calibrated by imaging a 1:1000 dilution of 5 mM Laurdan solution in DMSO at three different laser power settings (the same setting used for imaging the sample, as well as a setting 50$\%$ higher and 50$\%$ lower). Emission was collected by PMT detectors at 400-460 nm and at 470-530 nm. Detector gain and offset were held constant throughout the imaging. Analysis of the fluorescent images and determination of GP values were performed using the ImageJ macro developed previously \cite{54}. In addition, the shell of the bubble was segmented by taking pixel border equivalent to $\approx$1 $\mu$m around each bubble \cite{49}. Note that the fluorescence emission from the bubble membrane is required for calculation of membrane GP; NBs are too small for this purpose. For this reason, larger bubbles were chosen. Fig. 1a shows a schematic representation of how C-laurdan, PG, and Gly are assembled in the PL membrane. 
\subsection{Formulation of Nanobubbles (NBs)}
\begin{figure}
	\begin{center}	
		\includegraphics{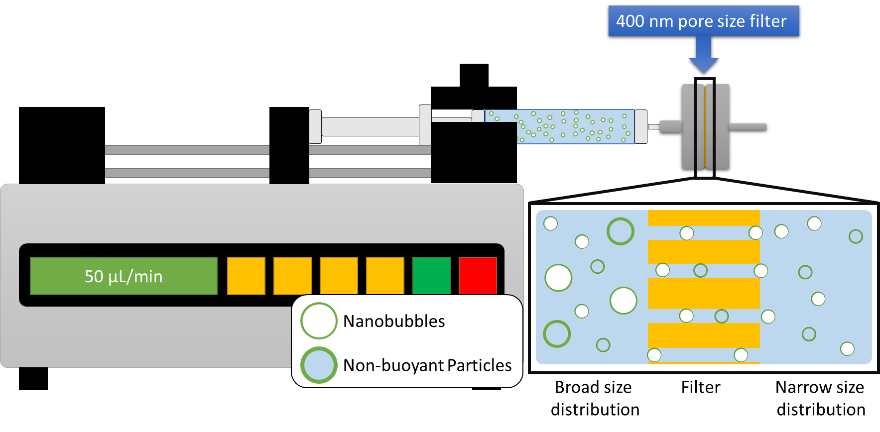}
		\caption{Schematic of filtration set-up to narrow down the size distribution of NB.}
	\end{center}
\end{figure}
As previously described \cite{52,53}, air was manually removed from lipid solutions in sealed vials using a 30 mL syringe and was replaced by $C_3F_8$ gas. and the phospholipid solution was activated by mechanical shaking with a VialMix shaker for 45s. Nanobubbles were isolated from the mixture of foam and microbubbles by centrifugation at 50 rcf for 5 mins with the headspace vial inverted, and 100 $\mu$L NB solution withdrawn from a fixed distance of 5 mm from the bottom with a 21G needle. To better highlight the effect of NB shell stiffness and eliminate the influence of size on the acoustic response the size distribution was narrowed via filtration through a 400-nm pore size filter, shown schematically in Fig. 2.  Isolation by differential centrifugation alone is insufficient to isolate NBs of a narrower size distribution brought about the low NB terminal velocity (i.e. calculated to be 22 nm/s for a 200-nm bubble) \cite{55}. The concentration and size distribution before and after filtration were characterized by resonant mass measurement \cite{52,53,56}.
\subsection{Acoustic measurements}
1 mL of NBs with narrow size distribution (5.0 x10$^8$ NBs/mL) was placed in an agarose phantom container for nonlinear ultrasound imaging. Nonlinear ultrasound imaging was carried out on an AplioXG SSA-790A clinical ultrasound imaging system (formerly Toshiba Medical Imaging Systems, now Hitachi Healthcare) with a 12 MHz center frequency linear array transducer (PLT-1204BT). Images were acquired in contrast harmonic imaging (CHI) mode with parameters set as: 65 dB dynamic range, 70 dB gain, receiving frequency 12 MHz and peak negative pressure 74 to 857 kPa. The agarose phantom was composed of 1.5 wt$\%$ agarose in Milli-Q water (resistivity of ~18 M$\omega$·cm) heated in a microwave until the agarose is dissolved. The hot agarose solution was then poured into a mold avoiding any trapped bubbles and cooled down to obtain phantom with the desired channel dimension as shown in Fig. 3. Intensity of the backscattered nonlinear ultrasound signal was determined using a pre-loaded quantification software (CHI-Q) setting the ROI to be around inside the channel as shown in Fig. 7 (top left image). The experiments were replicated three times. Enhancement was calculated by normalizing the measured backscattered ultrasound intensity of the NB solution with respect to the backscattered ultrasound intensity of the agarose phantom selected from an ROI at the same depth as the solution ROI.  
\begin{figure}
	\begin{center}	
		\scalebox{0.4}{\includegraphics{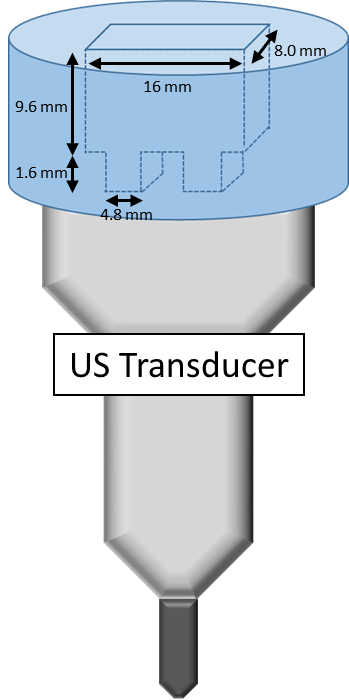}}
		\caption{Schematic of agarose phantom used for US imaging.}
	\end{center}
\end{figure}
\subsection{Numerical simulations}
\subsubsection{The bubble model}
The influence of the viscoelastic properties of the shell on NB dynamics were investigated using the modified Marmottant model \cite{18}. The Marmottant model was modified by Li et al., where the effects of shear thinning of the shell were added to the Marmottant model \cite{57,58}.  The model was recently used by Pellow et al. to investigate the NB dynamics \cite{20}.
The modified Marmottant model can be presented as:
\begin{linenomath*}
	\begin{equation}
	\begin{gathered}
	\rho \left(R\ddot{R}+\frac{3}{2}\dot{R}^2\right)=\\\left[P_0+\frac{2\sigma(R_0)}{R_0}\right](\frac{R}{R_0})^{-3k}\left(1-\frac{3k}{c}\dot{R}\right)- P_0-\frac{2\sigma(R)}{R}-\frac{4\mu_L\dot{R}}{R^2}-\frac{4k_s\dot{R}}{R^2}-P_a(t)
	\end{gathered}
	\end{equation}
\end{linenomath*}
In this equation, R is radius at time t, $R_0$ is the initial MB radius, $\dot{R}$ is the wall velocity of the bubble, $\ddot{R}$ is the wall acceleration,	$\rho{}$ is the liquid density (998 $\frac{kg}{m^3}$), c is the sound speed (1481 m/s), $P_0$ is the atmospheric pressure, k is the polytropic constant (1.068 for $C_3F_8$), $\sigma(R)$ is the surface tension at radius R, $\mu_L$ is the liquid viscosity (0.001 Pa.s), $k_s$ is the shell viscosity. The values in the parentheses are for pure water at 293 K. In this paper the gas inside the MB is $C_3F_8$ and water is the host media.  $P_a(t)$ is the amplitude of the acoustic excitation ($P_a(t)=P_a sin(2\pi f t)$) where $P_a$ and \textit{f} are the amplitude and frequency of the applied acoustic pressure. \\ 
The surface tension $\sigma(R)$ is a function of radius and is given by:
\begin{linenomath*}
	\begin{equation}
	\sigma(R)=
	\begin{dcases}
	0  \hspace{3cm}   if \hspace{1cm}  R \leq R_b\\
	\chi(\frac{R^2}{R_b^2}-1) \hspace{1.2cm} if\hspace{1cm}  R_b \leq R\leq R_{r}\\
	\sigma_{water} \hspace{2cm}  if\hspace{.5cm} Ruptured \hspace{.5cm}  R \geq R_r
	\end{dcases}
	\end{equation}
\end{linenomath*}
$\sigma_{water}$ is the water surface tension (0.072 N/m), $R_b=\frac{R_0}{\sqrt{1+\frac{\sigma(R_0)}{\chi}}}$ is the buckling radius, $R_r=R_b\sqrt{1+\frac{\sigma_{rupture}}{\chi}}$ is the rupture radius, and $\chi$ is the shell elasticity. Shear thinning of the shell is included in the Marmottant model using\cite{57,58}:
\begin{linenomath*}
	\begin{equation}
	k_s=\frac{4k_0}{1+\alpha \frac{|\dot{R}|}{R}};
	\end{equation}
\end{linenomath*}
where $k_0$ is the shell viscous parameter and $\alpha$  is the characteristic time constant associated with the shear rate. In this work $\alpha=0.75 \pm 0.25 \mu s$ which is in the range examined in \cite{20,57,58}.\\
$\sigma_{rupture}$ has been varied between 0.072 N/m for water to 1 N/m for different shells in the original work of Marmottant \cite{18}. When the bubble is compressed below its buckling radius, the effective surface tension on the bubble becomes zero. Above the buckling radius and below the break up radius the effective surface tension follows a linear elastic relationship. Above the rupture radius the effective surface tension on the bubble becomes equal to that of water. This is because the molecules of the shell will become farther apart leaving the bare gas exposed to water \cite{18}.\\
In this work, the frequency of the insonation is fixed at 6 MHz (the frequency used in the experiments), the excitation pressure amplitude is between 74-1249 kPa (pressure amplitude used in the experiments), the pulse duration is 2 cycles and the $R_0$ of the NBs is 100 nm (comparable to the mean diameter of 200 nm measured in the experiments).   
\subsubsection{Scattered pressure}
Oscillations of a bubble generate a scattered pressure ($P_{Sc}$) which can be calculated \cite{60}:
\begin{linenomath*}
	\begin{equation}
	P_{sc}=\rho\frac{R}{d}(R\ddot{R}+2\dot{R}^2)
	\end{equation}
\end{linenomath*}
where $d$ is the distance from the center of the bubble (and for simplicity is considered as 1m in this paper) \cite{61}.  The second harmonic component of the scattered pressure (at 12 MHz consistent with experiments)  was analyzed to compare the simulation results to the received signals in the experiments. In our numerical simulations shell elasticity was varied between 0.1 and 10 N/m, $\sigma(R_0)$   was varied between 0-0.072 N/m. The rupture surface tension was varied between that of water (0.072 N/m) and 1 N/m and shell viscosity was varied between $1*10^{-10}$ kg/s and $6*10^{-8}$ kg/s. 
\section{Results}
\subsection{Shell lipid packing order and stiffness}
\begin{figure}
	\begin{center}	
		\scalebox{0.5}{\includegraphics{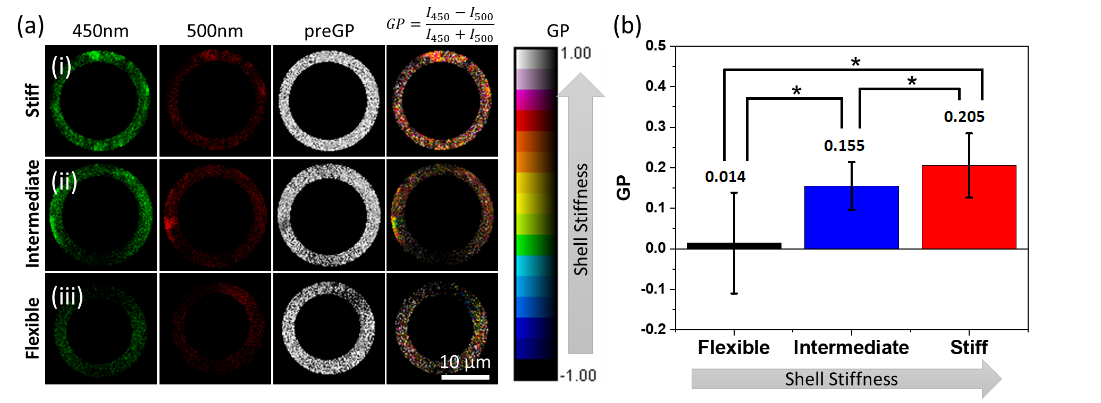}}
		\caption{ (a) Fluorescent images, pre-GP, and GP-images of shell membrane with different additives at 450- and 500-nm emission wavelength. (b) Comparison of average GP for bubbles with different shell stiffness (n = 55 for each bubble type).}
	\end{center}
\end{figure}
Incorporation of Gly (20$\%$ v/v) dehydrates the PL membrane, which increases the PL packing order as shown schematically in Fig. 1. The increase in PL packing order causes the C-laurdan to emit higher intensity light at 450-nm compared to 500-nm since it is surrounded by a less polar environment (Fig. 4a). The GP value (formula indicated in Fig. 4a) for each pixel was calculated and averaged throughout the whole bubble shell. The average GP value for PL membrane with Gly was determined to be 0.205 (Fig. 4b). On the other hand, incorporation of PG (20$\%$ v/v) in the PL membrane increases the distance between the PL molecules thereby letting more water surround C-laurdan (Fig. 1). The emission of C-Laurdan at 450-nm has similar intensity as compared to at 500-nm (Fig. 4a, iii). The mean GP for the PL membrane with PG was calculated to be 0.014 (Fig. 4b), which is less than the mean GP for PL with Gly. Therefore, incorporation of PG results in an increase in PL disorder and consequent decrease in membrane stiffness. The measurement of GP for the PL membrane with PG or Gly provides an additional confirmation that incorporation of Gly increases the PL packing order, consistent with what Terakosolphan et al. and Pocivavsek et al. have reported \cite{41,62}. Incorporation of both Gly (10$\%$ v/v) and PG (10$\%$ v/v) resulted in an average GP value of 0.155 (Fig. 4b) that is in-between the GP values for PL with Gly and PL with PG. This suggests that the C-laurdan in the membrane is surrounded by relatively polar environment in some areas and relatively non-polar environment in other areas as schematically shown in Fig. 1a. The difference in PL packing order through addition of different membrane additive is expected to have significant impact on the shell stiffness, and subsequently on how NBs interact with US. 
\subsection{Size isolated NBs}
\begin{figure}
	\begin{center}	
		\scalebox{0.4}{\includegraphics{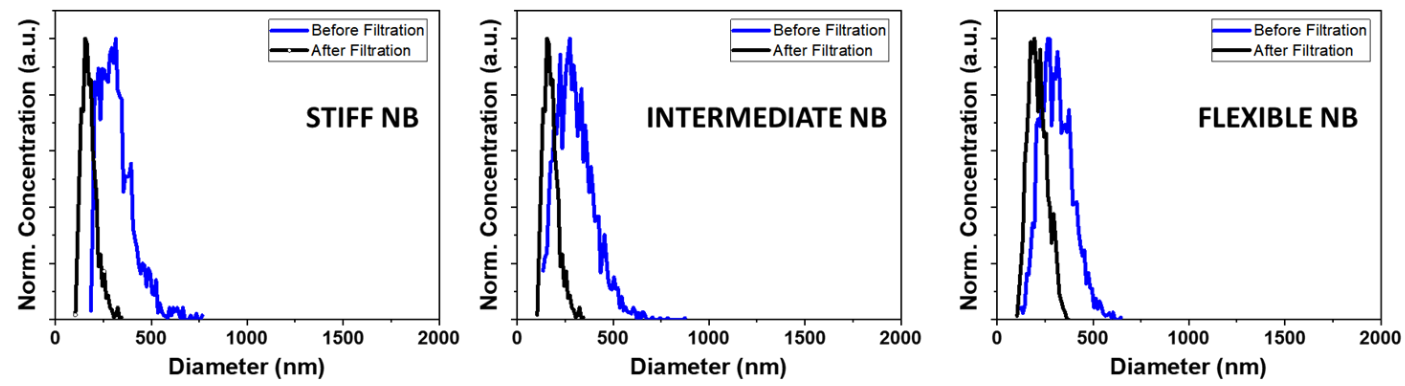}}
		\caption{ Size distribution and concentration of NBs, characterized by a resonant mass measurement, of each type before and after filtration through a 400-nm pore membrane filter.}
	\end{center}
\end{figure}
After centrifugation, the size distribution and concentration were determined using a resonant mass measurement system before and after filtration (Fig. 5 , Table 1) as previously described \cite{63}. Although no bubbles larger than 1 $\mu$m can be observed in the unfiltered population, the size of NBs broadly ranged from 100 nm to about 800 nm with a mean diameter of 310 $\pm$ 10 nm for Flexible NB, 301 $\pm$ 9 nm for Intermediate NB, and 318 $\pm$ 11 nm for Stiff NB. The broad size distribution of the population hinders the accurate study of the shell viscoelasticity on the NB dynamics.   The size distribution of filtered NBs (Fig. 5, black trace) show a mean size of 213 $\pm$ 5 nm for Flexible NB, 176 $\pm$ 3 nm for Intermediate NB, and 178 $\pm$ 5 nm for Stiff NB nm (Table 1), with no NBs larger than 400 nm observed for all groups. NB solutions were of different concentration after filtration but were adjusted to an approximate number density of 5.0 x10$^8$ NBs/mL by addition of PBS for subsequent US studies. 
\begin{table}[]
	\begin{center}
		\begin{tabular}{|l|l|l|l|l|l|l|}
			\hline
			& \multicolumn{3}{l|}{Before} & \multicolumn{3}{l|}{After} \\ \hline
			& min   & max  & mean         & min   & max  & mean        \\ \hline
			Flexible NB     & 125   & 645  & 310 $\pm$ 10 nm  & 105   & 375  & 213 $\pm$ 5 nm  \\ \hline
			Intermediate NB & 135   & 875  & 301 $\pm$ 9 nm   & 105   & 345  & 176 $\pm$ 3 nm  \\ \hline
			Stiff NB        & 185   & 765  & 318 $\pm$ 11 nm  & 105   & 325  & 178 $\pm$ 5 nm  \\ \hline
		\end{tabular}
		\caption{Min, max, and mean size before and after filtration of NBs with different additives. The size distributions are shown in Figure 5.}
	\end{center}
\end{table}

\subsection{Acoustic signals from NBs}
\begin{figure}
	\begin{center}	
		\scalebox{0.35}{\includegraphics{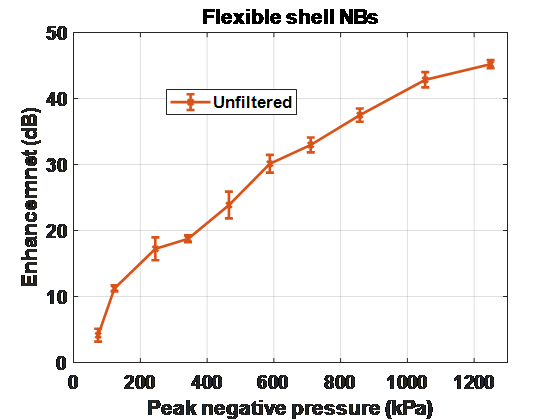}}\scalebox{0.35}{\includegraphics{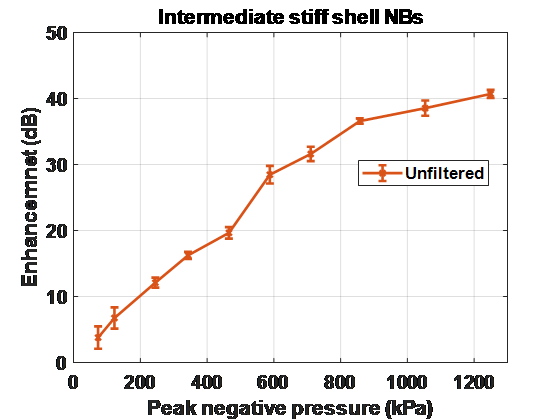}}\scalebox{0.35}{\includegraphics{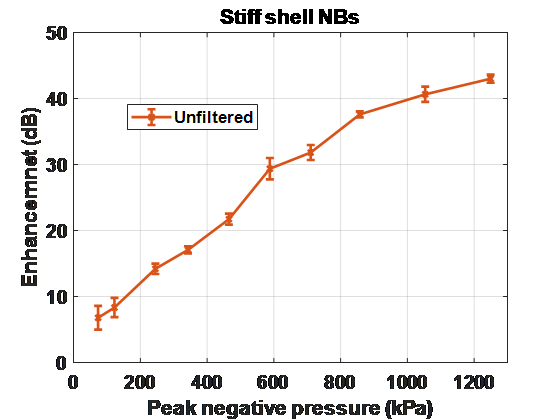}}\\
		(a) \hspace{5cm} (b) \hspace{4cm} (c)
		\caption{Contrast enhancement of polydisperse NB solution with (a) Flexible , (b) Intermediate, and (c) Stiff shells, relative to the agarose phantom for different peak negative pressures (PNP).  Error bars are standard deviation of three independent replicates.}
	\end{center}
\end{figure}
Results of the acoustic measurements of the unfiltered polydisperse populations are shown in Fig. 6. To quantify the nonlinear signal from the NB solution, the raw US echo power was averaged over the region of interest and the enhancement was calculated relative to the signal from the surrounding agarose phantom at the same depth. Fig. 6 shows that, there is no clear difference between the received signals from the three populations, most likely due to the polydisperse nature of the NB solutions that masks the shell effects.\\
Fig. 7 shows the comparison between the 2nd harmonic contrast enhanced images of the filtered monodisperse  NB solutions. There is a clear difference between the echogenicity of the three NB populations. This is witnessed by a sudden increase in the contrast enhancement of the Flexible NBs at 245 kPa, followed by the sudden enhancement at  465 kPa and 588 kPa for the intermediate and stiff shell NBs respectively. The flexible NB solution undergoes another sudden enhancement at 857 kPa followed by loss of echogenicity at 1053 kPa (possibly due to NB destruction, and the mechanism is explored later in the results section).\\
Fig. 8 shows the enhancement as a function of pressure for all three filtered formulations. To better identify the pressure threshold for the signal the sudden amplification, we also plot the slope of the contrast enhancement as a function of pressure. To plot these graphs, the raw US echo power was averaged over the region of interest (white dashed square in Fig.  8) and the enhancement was calculated relative to the signal from the surrounding agarose phantom at the same depth. Compared to Fig. 6, a substantial difference in enhancement was observed for the US signal from NBs before and after filtration for all formulations. Narrowing of the size distribution by filtration (black traces in Fig. 5) yielded clear activation pressure thresholds for all bubble types. This threshold was not detectable for the unfiltered NBs (blue traces (Fig. 5)).\\ 
For Flexible NBs, the PNP was varied between 74 to 1250 kPa as shown in Fig. 8a. There is no detectable nonlinear activity at PNP between 74-123 kPa, (MI = 0.03-0.05). Increasing the PNP to 245 kPa results in a 14 dB increase in enhancement. A significant increase in enhancement occurred when the PNP was increased from 123 kPa to 245 kPa, with a slope of 0.11 dB/kPa (Fig. 8d). The absence of detectable signal from Filtered Flexible NBs at low PNP implies that these NBs oscillation at this pressure is very weak, thus the signal generated is not within the detectable range of the US transducer.  This behavior is unlike the polydisperse NB solution where there is no observable pressure threshold ($P_t$) for the Unfiltered Flexible NB solution.  Further increase in PNP results in another sudden enhancement in pressure amplitude at 710-857 kPa with a slope of 0.12 dB/KPa.\\
\begin{figure}
	\begin{center}	
		\scalebox{0.4}{\includegraphics{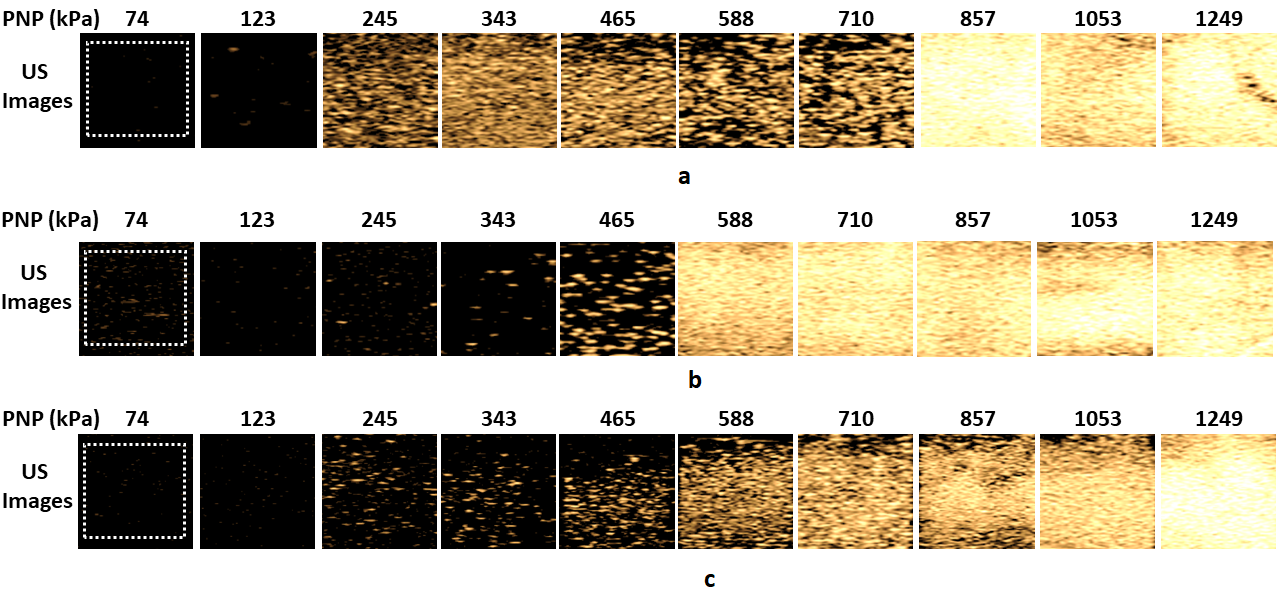}}
		\caption{ Representative US CHI mode contrast image of solution of filtered monodisperse (a) Flexible, (b) Intermediate and (c) Stiff shell NBs for PNP = 74 – 1250 kPa. }
	\end{center}
\end{figure}
For NBs containing both Gly and PG (Intermediate NBs), the presence of the two membrane additives results in a membrane stiffness between that of the membrane with PG and membrane with Gly, as confirmed by intermediate GP value in Fig. 4b \cite{52}.  Similar to the solution of the Filtered Flexible NB, there is a negligible detectable nonlinear activity when Filtered Intermediate NBs were exposed to a PNP below 343 kPa, as shown in Figs. 8b and 8e. As soon as the PNP is increase above 465 kPa, the signal is enhanced suddenly with a slop of 0.14 dB/kPa at 465 kPa (Fig.8e). Further increase in the PNP to 1250 kPa resulted in a steady increase in the enhancement.\\
For the Filtered monodisperse Stiff NBs, a steady increase in enhancement was measured between 343 to 465 kPa. Further increases in PNP to 588 kPa resulted in a substantial increase in brightness that continued to increase up to a PNP of 710 kPa. Analysis of the raw echo power and enhancement as a function of PNP reveals that a threshold pressure for a sudden amplification ($P_t$) exists between PNPs of 588 to 710 kPa with a slope of 0.08 dB/kPa. Such a transition region is not observed with the solution of unfiltered NBs, likely due to the effect of a broad NB size distribution on the scattering. 
The first $P_t$ for Filtered Flexible NB solution occurs at a lower pressure range (123-245 kPa) as compared to the Intermediate NB solution (465-588 kPa) and Filtered Stiff NB solution (588-710 kPa).  Moreover, only the Flexible NB solution exhibits the second amplification $P_t$ at 710-857 kPa. 
\begin{figure*}
	\begin{center}
		\scalebox{0.3}{\includegraphics{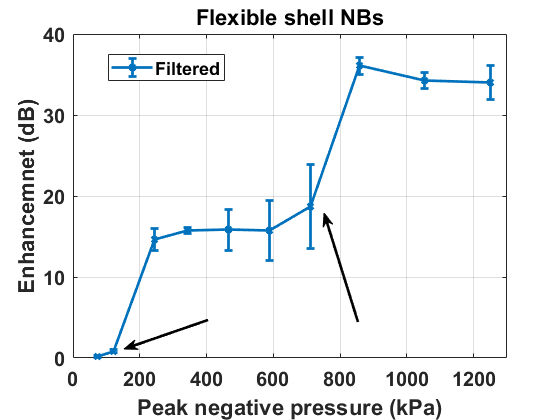}}  \scalebox{0.3}{\includegraphics{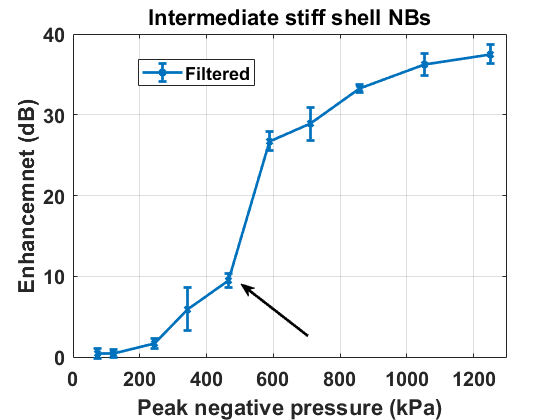}}\scalebox{0.3}{\includegraphics{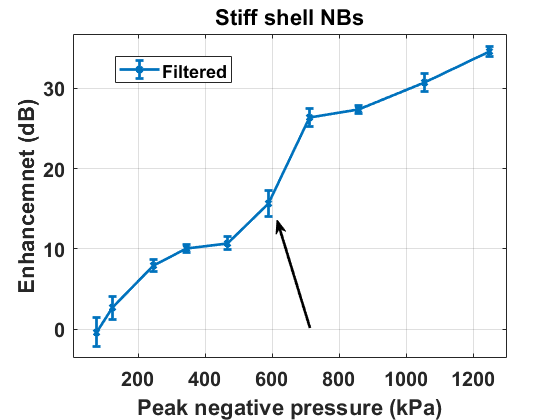}}\\
		(a) \hspace{4cm} (b) \hspace{4cm} (c)\\
		\scalebox{0.3}{\includegraphics{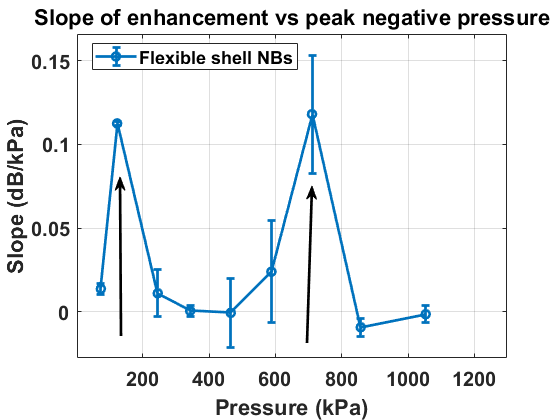}}  \scalebox{0.3}{\includegraphics{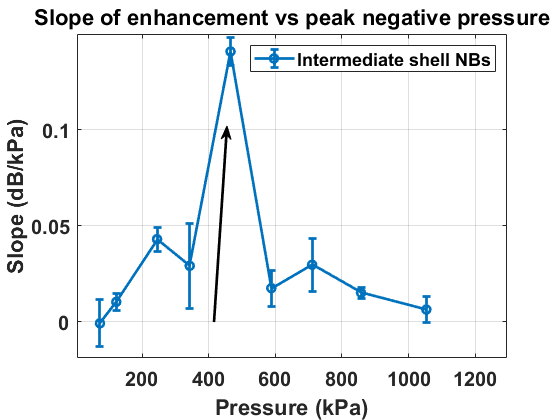}}\scalebox{0.3}{\includegraphics{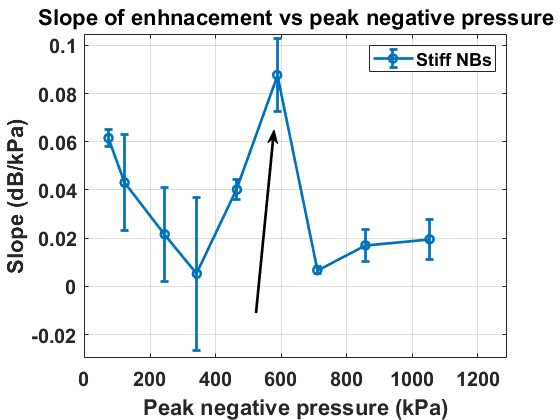}}\\
		(e) \hspace{4cm} (f) \hspace{4cm} (g)
		\caption{Contrast enhancement of filtered monodisperse NB solution with (a) flexible, (b) Intermediate, and (c) Stiff shells relative to the agarose phantom for different peak negative pressures (PNP). The slope of the contrast enhancement with respect to peak negative pressure for (d) Flexible, (e) Intermediate and (f) Stiff shell NBs. Arrows mark the pressure threshold ($P_t$) of sudden signal enhancement. Error bars are standard deviation of three independent replicates. }
	\end{center}
\end{figure*} 
These results suggest that there is a strong correlation between the $P_t$ of different NB formulations and their relative shell stiffness as quantified by their average GP. Plotting the midpoint of the range of pressure values $P_t$ vs average GP (Fig.  9) reveals a linear dependence with an intercept of -0.06 $\pm$ 1.84*10$^{-4}$ and a slope of 0.041 $\pm$ 3.41*10$^{-5}$. This shows that there is a strong correspondence between NB shell stiffness and their nonlinear behavior under US.  
\begin{figure}
	\begin{center}	
		\scalebox{0.4}{\includegraphics{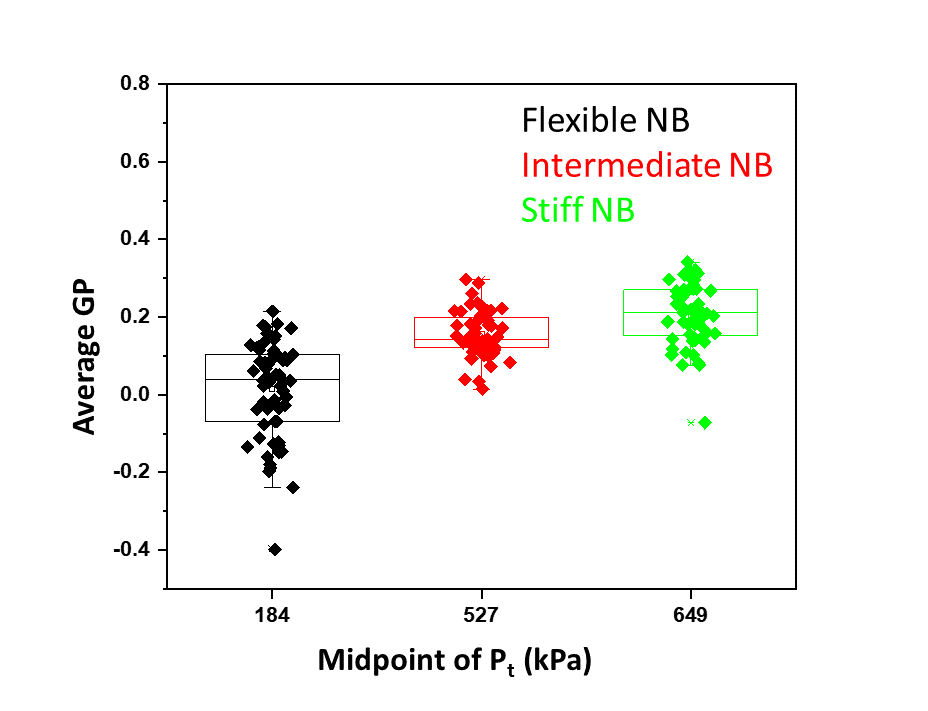}}
		\caption{ Correlation between the midpoint of the range of pressure values $P_t$ of NB of different shell stiffness and the average GP of its shell. }
	\end{center}
\end{figure}
\renewcommand{\floatpagefraction}{.8}
\begin{figure*}
	\begin{center}
		\scalebox{0.4}{\includegraphics{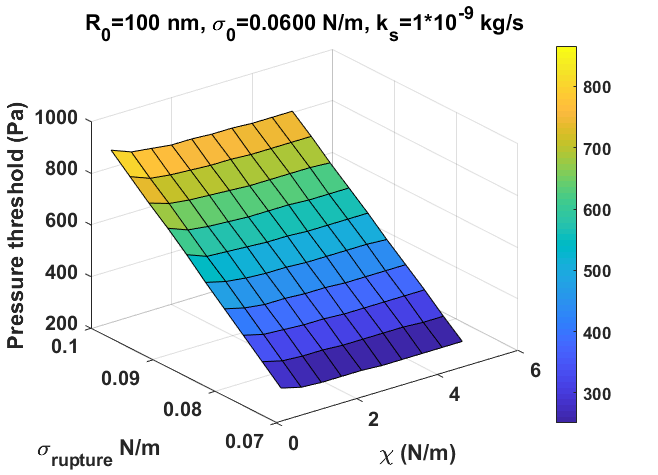}}  \scalebox{0.4}{\includegraphics{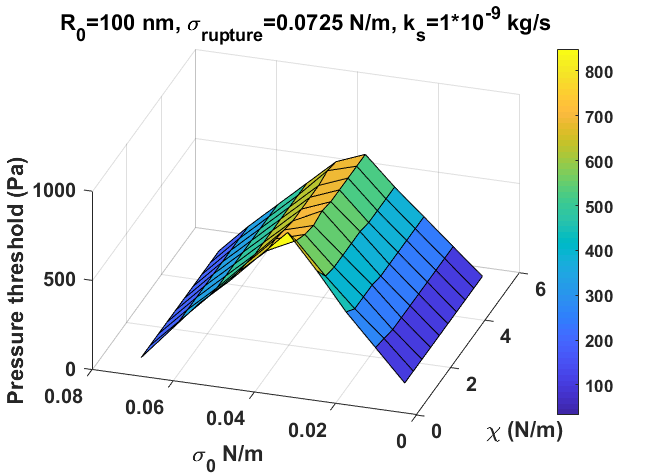}}\\
		(a) \hspace{6cm} (b)\\
		\scalebox{0.4}{\includegraphics{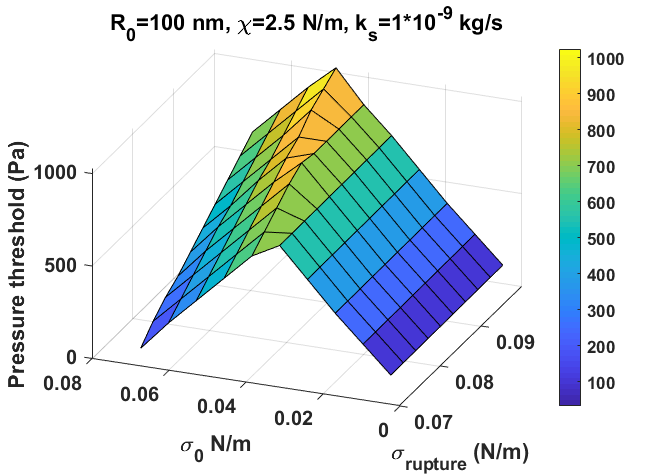}}  \scalebox{0.4}{\includegraphics{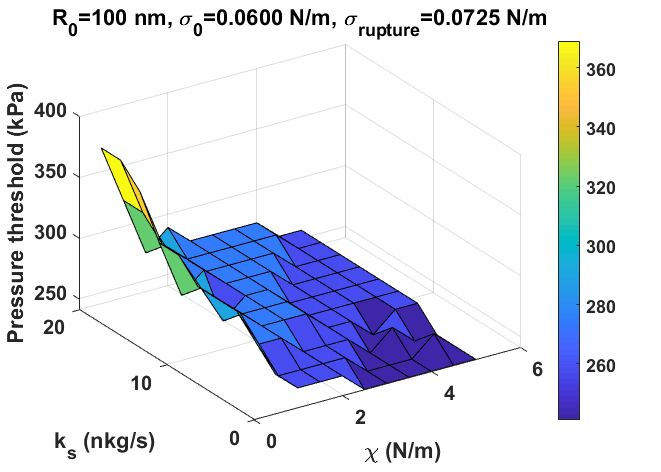}}\\
		(e) \hspace{6cm} (f) 
		\caption{Pressure threshold of the sudden enhancement in the 2nd SuH (Z-axis) of a NB with $R_0$ =100 nm as a function of : a) $\chi-\sigma_{rupture}$, b)$\chi-\sigma(R_0)$ , c) $\sigma_{rupture}-\sigma(R_0)$ and d) $\chi-k_s$.}
	\end{center}
\end{figure*} 
\renewcommand{\floatpagefraction}{.8}
\begin{figure*}
	\begin{center}
		\scalebox{0.35}{\includegraphics{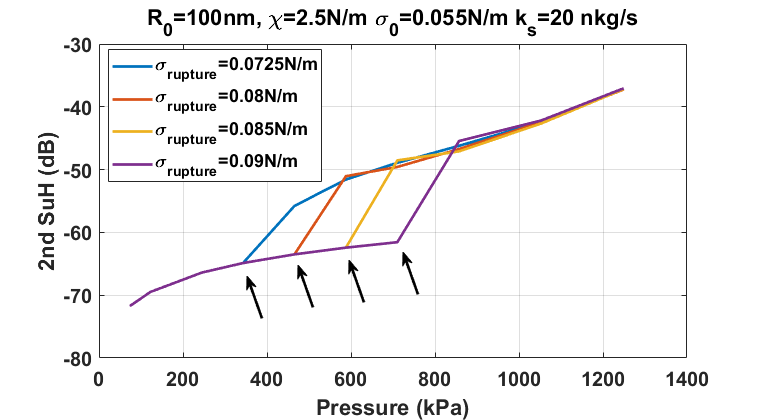}}  \scalebox{0.35}{\includegraphics{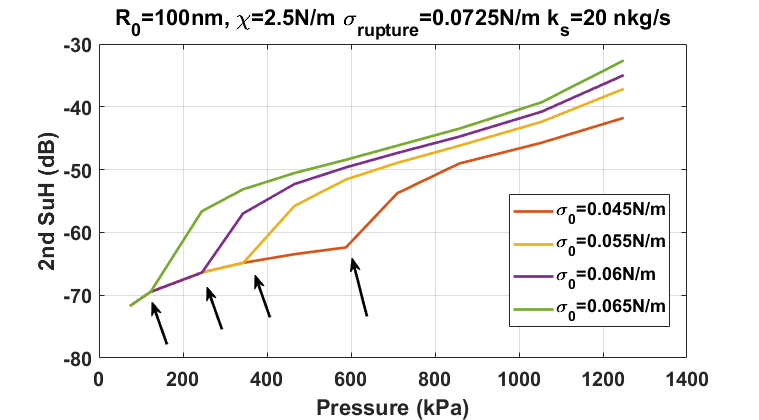}}\\
		(a) \hspace{6cm} (b)\\
		\scalebox{0.35}{\includegraphics{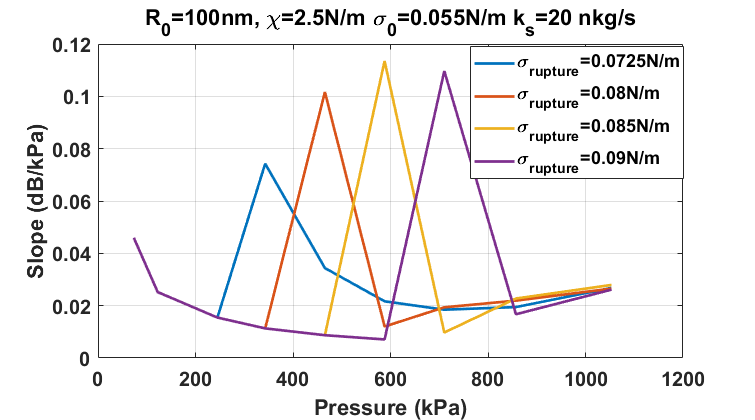}}  \scalebox{0.35}{\includegraphics{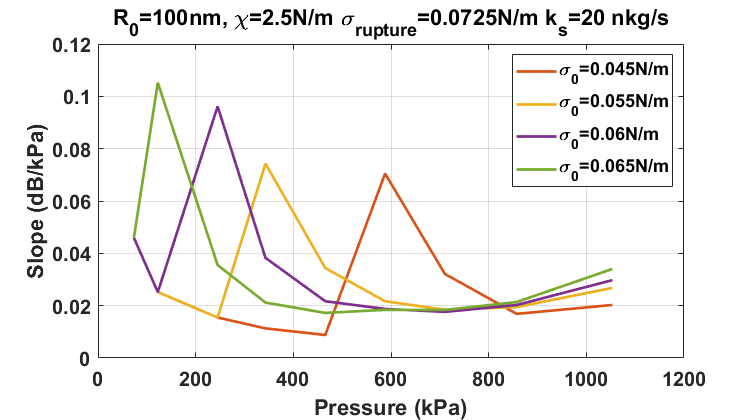}}\\
		(e) \hspace{6cm} (f) 
		\caption{The 2nd SuH amplitude of the scattered pressure as a function of excitation pressure for a NB with $R_0$=100nm, $k_s$=20 nkg/s and $\chi$=2.5 N/m : a)  for different $\sigma_{rupture}$ when $\sigma(R_0)$=0.055 N/m, b) for different $\sigma(R_0)$ when $\sigma_{rupture}$=0.0725 N/m. The corresponding slope of the 2nd SuH enhancement as a function of the excitation pressure.}
	\end{center}
\end{figure*} 
\subsection{Numerical simulations}
In order to investigate the mechanism behind the observed changes in the $P_t$ for different NBs we ran simulations over a large range of parameters and visualized the results of the 2nd harmonic component (2nd SuH \cite{59}) of the scattered pressure and the slope of the 2nd SuH as a as a function of excitation pressure amplitude.  In this section we show the effect of the different shell parameters ($R_0$ ),  $\sigma_{rupture}$, $\chi$,  and $k_s$ on the pressure threshold of the enhancement in  2nd SuH. Next, the shell parameters values that best fit the slope vs pressure curve in each case will be calculated. The reason the slope vs the pressure was chosen as the fitting curve is to minimize the influence of the parameters that lead to quantitative differences between the modeled 2nd SuH and the enhancement amplitude in experiments.  As the slope curve is relative to before and after the enhancement, its magnitude should be better matched between the experiments and the numerical simulations. This is because the different contributing factors may be canceled due to the relative nature of the slope curves, leaving only the enhancement difference. 
\renewcommand{\floatpagefraction}{.8}
\begin{figure*}
	\begin{center}
		\scalebox{0.35}{\includegraphics{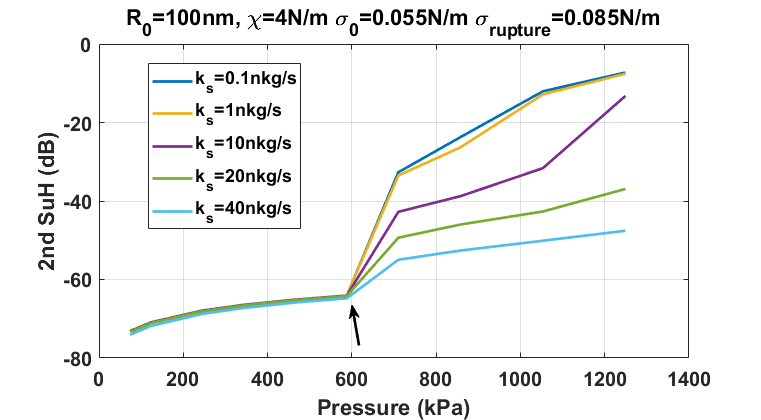}} \scalebox{0.35}{\includegraphics{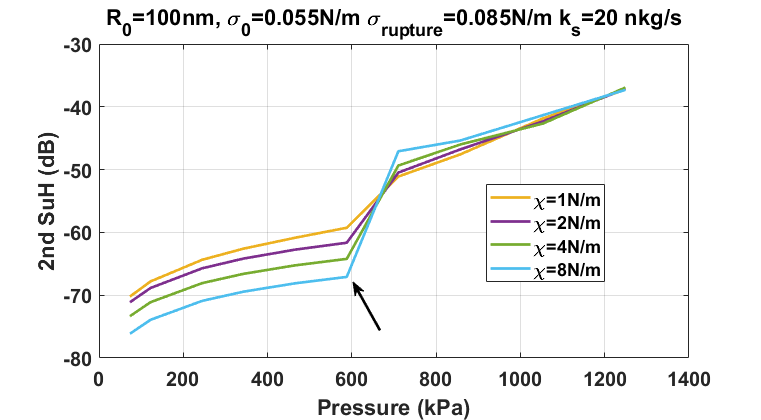}}\\
		(a) \hspace{6cm} (b)\\
		\scalebox{0.35}{\includegraphics{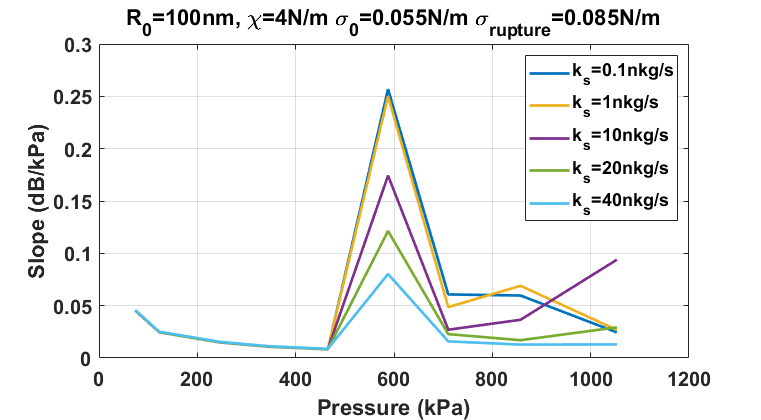}} 	\scalebox{0.35}{\includegraphics{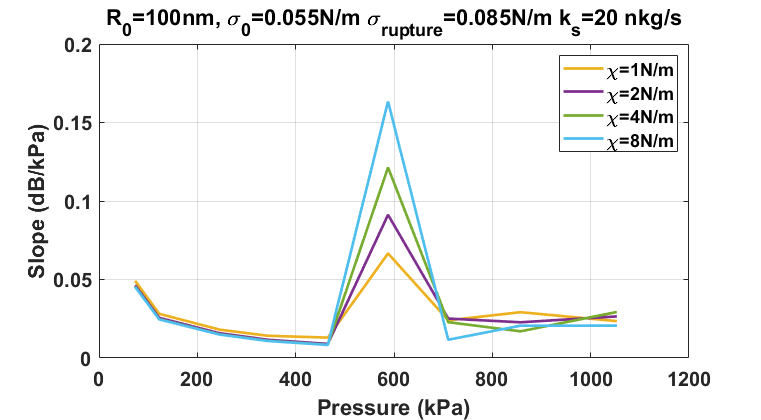}} \\
		(e) \hspace{6cm} (f) 
		\caption{The 2nd SuH amplitude of the scattered pressure as a function of excitation pressure for a NB with $R_0$=100nm, $\sigma(R_0)$=0.055 N/m and $\sigma_{rupture}$= 0.085 N/m: a)  for different $k_s$  when $\chi$=4 N/m b) for different $\chi$ when $k_s$=20 nkg/s. (c) and (d) The corresponding slope of the 2nd SuH enhancement as a function of the excitation pressure.}
	\end{center}
\end{figure*} 
\begin{figure*}
	\begin{center}
		\scalebox{0.3}{\includegraphics{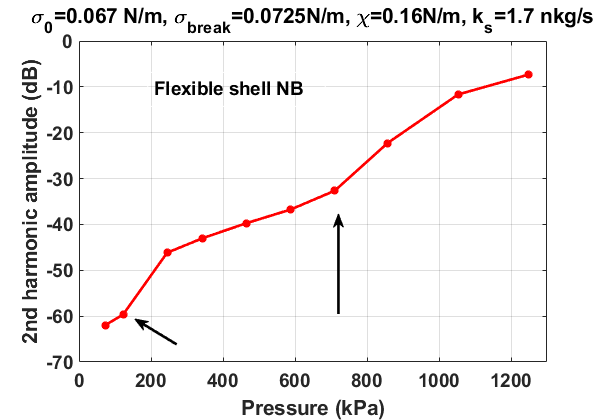}}  \scalebox{0.3}{\includegraphics{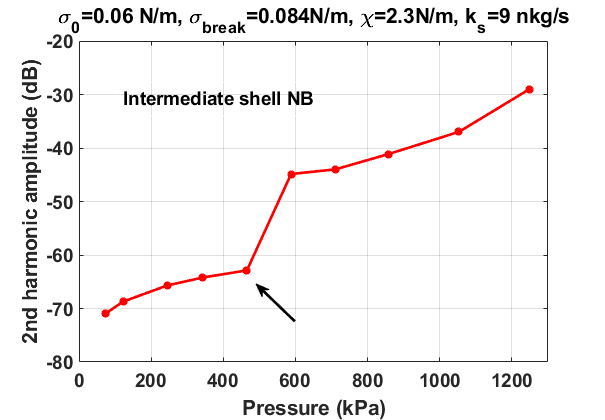}}\scalebox{0.3}{\includegraphics{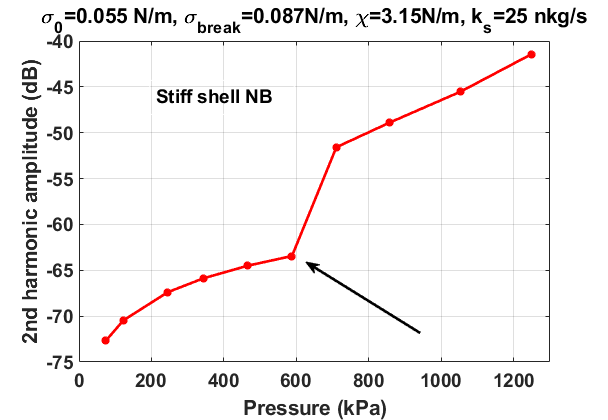}}\\
		(a) \hspace{4cm} (b) \hspace{4cm} (c)\\
		\scalebox{0.3}{\includegraphics{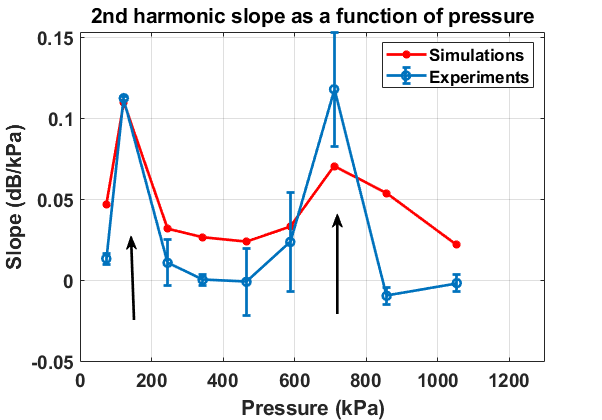}}  \scalebox{0.3}{\includegraphics{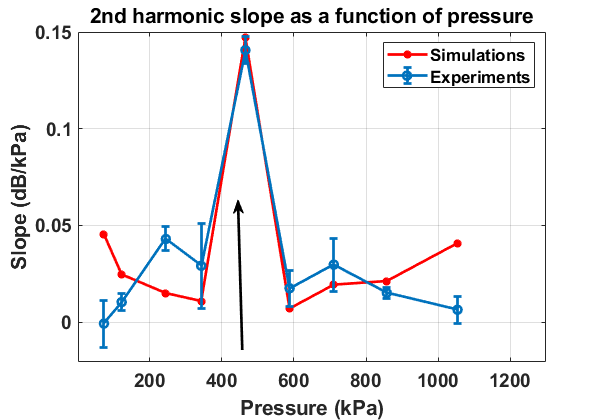}}\scalebox{0.3}{\includegraphics{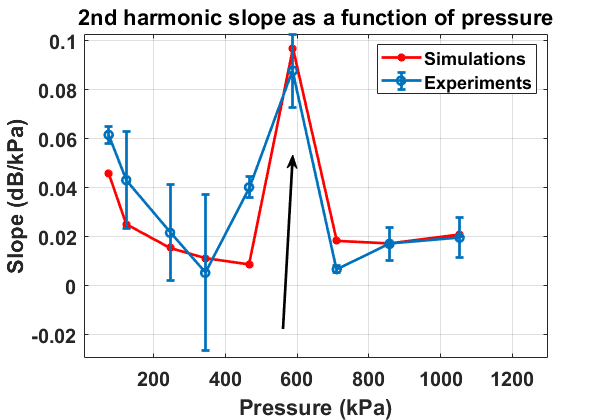}}\\
		(e) \hspace{4cm} (f) \hspace{4cm} (g)
		\caption{The 2nd SuH frequency component of the numerically calculated scattered pressure of a NB with $R_0$=100nm with: (a) Flexible, (b) Intermediate, and (c) Stiff shells. Comparison between the slope of the contrast enhancement with respect to the excitation pressure amplitude between numerical simulations and experiments for: (d) Flexible, (e) Intermediate and (f) Stiff shell NBs. Arrows mark the pressure thresholds (Pt) of the sudden signal enhancement. Error bars are standard deviation of the three independent replicates.}
	\end{center}
\end{figure*} 
\subsubsection{Influence of the shell properties on the threshold behavior}
Figure 10 shows the pressure threshold for the sudden amplification of the 2nd SuH as a function of the shell parameters. The changes in the value of shell elasticity have no (or minimal) effect on the pressure threshold ($P_t$) of the amplification (Figs. 10a-10b). However, changes in $\sigma(R_0)$  and  $\sigma_{rupture}$ have significant influence on the $P_t$. For a constant $\sigma(R_0)$, $P_t$ increases with increasing $\sigma_{rupture}$ (Fig. 10a). For a constant $\sigma_{rupture}$  (using the water surface tension of 0.0725 N/m), there are two scenarios for the dependence of the $P_t$. $P_t$ increases with increasing $\sigma(R_0)$ until it reaches $\sigma_{rupture}/2$, beyond which $P_t$ decreases with increasing $\sigma(R_0)$ (Fig. 10 b). Fig. 10c shows the $P_t$ as a function of   $\sigma(R_0)$  and  $\sigma_{rupture}$ for constant $k_s$ and $\chi$. For  $\sigma(R_0)<\sigma_{rupture}/2$, increasing $\sigma_{rupture}$ has no effect on the $P_t$, however, for $\sigma(R_0)<\sigma_{rupture}/2$, increasing $\sigma_{rupture}$ increases the $P_t$, with the highest rate of increase for $\sigma(R_0)=\sigma_{rupture}/2$.  Higher $k_s$, may increase the $P_t$ (Fig. 10d), however, the influence of the $k_s$ on the $P_t$ is orders of magnitude smaller than the influence of $\sigma_{rupture}$ and $\sigma(R_0)$. $k_s$ has a stronger effect of the $P_t$ for a NB with smaller $\chi$.\\ 
Figure 11 shows the influence of the $\sigma_{rupture}$ and $\sigma(R_0)$ on the $P_t$ and the slope of the 2nd SuH enhancement. For a given initial surface tension above 0.036 N/m, $P_t$ (Fig. 11a) increases with increasing $\sigma_{rupture}$ with no apparent relation between the slope of enhancement (in dB/kPa) and $\sigma_{rupture}$ (Fig. 11c). For a given $\sigma_{rupture}$ and for $\sigma(R_0)>0.036 N/m$, $P_t$  (Fig. 11b) and slope of enhancement (Fig. 11d) decrease with increasing $\sigma(R_0)$.\\
Figure 12 shows the influence of $\chi$ and $k_s$ on the $P_t$ and the slope of 2nd SuH enhancement. Changes in $\chi$ and $k_s$ do not have any effect on the $P_t$ (Fig. 12a-b). However, the slope of 2nd SuH enhancement decreases with increasing $k_s$ (Fig. 12c) and increases with increasing $\chi$ (Fig. 12d). Using the information gained by analyzing Figs. 10-12, numerical simulations were performed for different values of the $\sigma_{rupture}$, $\sigma(R_0)$,  $\chi$ and $k_s$ (given in methods section) and the results of the best fit to the experimental slope curves are presented in Fig. 13. There is an excellent agreement between the numerical simulations and the experiments for a) the pressure threshold of enhancement and b) the slope of the enhancements. Numerical simulations predict the two experimentally observed pressure thresholds for the enhancement of the signal from the flexible NB solutions at 125 kPa and 857 kPa (Fig. 13a). There is also a very good agreement between the numerical and experimental slope curves (Fig. 13d).  In agreement with experiments the simulations predict the $P_t$ of 465kPa (Fig. 13b) and 588kPa (Fig. 13c) for the intermediate and the stiff shell NBs. The corresponding numerically calculated slope curves have qualitative and quantitative agreement with experimentally measured curves (Figs. 13e-f). In agreement with the GP measurements, numerical results predict the smallest shell elasticity for the flexible NBs ($\chi$=0.16 N/m) and medium elasticity of $\chi$=2.3 N/m for the intermediate NBs and the highest elasticity of  $\chi$=3.15 N/m for the stiff NBs. Moreover, it is numerically predicted that the addition of Gly increases the surface tension for rupture from 0.0725 N/m for the flexible NBs to 0.084 N/m for the intermediate NBs, and 0.087 N/m for the stiff NBs.  Addition of Gly is also accompanied by a reduction in the initial surface tension from 0.067 N/m for flexible NBs to 0.06 N/m for the intermediate NBs to 0.055 N/m for the stiff NBs. As expected, due to higher viscosity of Gly, addition of Gly also increases the viscosity of the shell from 1.7 nkg/s for the flexible NBs to 9 nkg/s for the intermediate NBs and 25 nkg/s for the stiff NBs. 
\subsection{Examination of the radial oscillations before and at the pressure threshold of signal enhancement}
Figure 14 shows the radial oscillations as a function of time for the Flexible NB in Fig. 13a. For pressure amplitudes below the first $P_t$, the radial oscillation amplitudes are small. The maximum oscillation amplitude is below $R_r=1.012R_0$. When $P_a=P_t$, radial oscillations grow above $R_r$ and as soon as the shell ruptures (marked in Fig. 14a), the rate of the growth of radial oscillations increases and the bubble expands very fast. This is the point where the first signal enhancement shown in Figs.  7a, 8a and 13a occur. In the collapse phase the bubble collapses rapidly but the speed of collapse decreases significantly after the radial oscillation is less than $R_r$; this is due the fact that the shell reseals and resists the fast collapse. At the higher pressure of $P_a=857 kPa$ a second enhancement occurs (Figs. 7a, 8a and 13a). In Fig. 7a, a loss of echogenicity takes place for $P_a\geq857 kPa$. The radial oscillations in Fig. 14b show that the maximum oscillation amplitude increases beyond the minimum reported threshold of bubble destruction ($R/R_0=2$ \cite{BBB,DDD}). This threshold is marked in Fig. 14b. In the collapse phase, due to the higher wall velocities (compared to the first $P_t$) the shell resistance after reseal is not strong enough to withstand the fast wall velocity, thus the NB collapses to a size close to half of its initial size and thus it is likely to undergo fragmentation. This can be one possible reason for the loss of echogenicty seen in Fig. 7a for pressures more than 857kPa. Moreover, since NB destruction is not modeled in the simulations, this can be a possible reason for the discrepancy between the simulations and measurements above the second $P_t$ (Fig. 13c). One discrepancy is seen in the peak of the slope graph at the 2nd $P_t$ in Fig. 13e. The next discrepancy occurs in the amplitude graph (Fig. 13a) above the second $P_t$. When the experimental amplitude graphs (Figs. 8a-c) are examined, the maximum enhancements are 36, 37.5 and 34.5 dB for flexible, intermediate and stiff NBs respectively. In the case of the flexible NBs, the maximum enhancement takes place at $P_t= 857kPa$, beyond which the enhancement decreases slightly with increasing pressure. In the case of intermediate and stiff NBs the maximum occurs at the highest examined pressure of 1249 kPa. Examining the 2nd harmonic amplitude in Figs. 13a-c, shows that this trend does not hold above the second pressure threshold of 857 kPa in the case of the flexible NBs. The maximum 2nd harmonic amplitude for intermediate and stiff NBs are -30 and -41 dB respectively. The maximum 2nd harmonic amplitude for the flexible NB is -32 dB right after the 2nd $P_t$ which shows a good agreement between trend observed in experiments and numerical simulations. However, beyond this point the trend does not hold and numerical simulations predict a constant growth in the 2nd SuH amplitude which at $P_a=1249 kPa$ reaches to -7 dB. This discrepancy beyond the second $P_t$ may be another indicator that the NBs undergo destruction which was not taken into account in the simulations.  
\begin{figure*}
	\begin{center}
		\scalebox{0.4}{\includegraphics{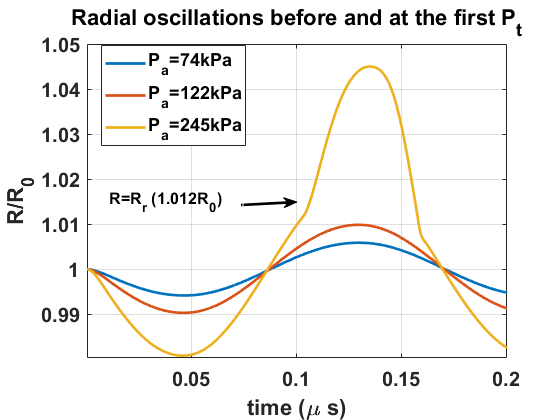}}  \scalebox{0.4}{\includegraphics{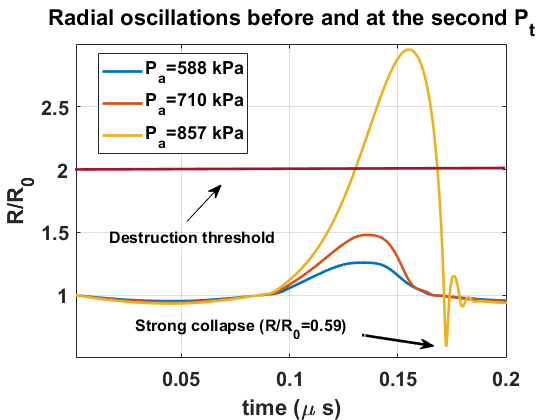}}\\
		(a) \hspace{6cm} (b)\\
		\caption{simulated radius versus time for the case of the flexible NB for: (a) radial oscillations before and at the first $P_t=245 kPa$ and (b) radial oscillations before and at the second $P_t=857 kPa$.}
	\end{center}
\end{figure*} 
\section{Discussion}
In this study, the shell viscoelastic properties were modified by the introduction of membrane propylene glycol (PG) as a membrane softener and glycerol (Gly) as a membrane stiffener. A two-photon microscopy technique through a polarity-sensitive fluorescent dye, C-Laurdan, was utilized to gain insights on the effect of membrane additives to the membrane structure. PG and Gly affect the structure and properties of the membrane of PL-stabilized UCAs, and therefore UCA response to an US field.
The solutions were sonicated with US pulses with 6 MHz center frequency and a PNP range of 100-857kPa. The Filtered NBs (200 nm mean diameter, narrow size distribution) exhibited a threshold behavior with increasing PNP. Above a pressure threshold the echogenicity of the 2nd harmonic contrast-mode CHI images of NBs increased suddenly. The pressure threshold for signal amplification increased with shell stiffness. Rapid enhancement of the second harmonic observed for PNP ranges of 123-245 kPa for the flexible membrane, 465-588 kPa for the intermediate membrane, and 588-710 kPa for the stiff membrane. The difference in the amplitude of the excitation pressure for threshold behavior may be explained by the shell composition properties with changes in elasticity, shell rupture threshold, initial surface tension and viscosity.\\  
\textbf{1-Increased elasticity}\\
The significant difference in measured GP (e.g. 0.205 for stiff shells, 0.155 for intermediate and 0.014 for flexible shells) reflects how Gly and PG interact with the PL membrane. Large parameter numerical simulations showed that the changes in the elasticity and shell viscosity do not have a significant influence on the pressure threshold, however, they largely affect the slope of the 2nd harmonic amplitude as a function of pressure.  The slope curves were used to fit the numerical simulations to the experimental measurements as they have information on both the pressure threshold and growth rate of the 2nd harmonic as a function of the excitation pressure. Moreover, due to the relative nature of the slope curves a good quantitative agreement between experiments and numerical simulations was achieved. According to numerical simulations, addition of Gly leads to an increase in shell elasticity  from 0.15 N/m for the flexible NBs to 2.3 N/m for Intermediate NBs and 3.15 N/m for the Stiff NBs.  Comparing the ratio of the predicted elasticities to the ratio of GP values also show a good correlation between experiments and numerical simulations. The ratio of the predicted elasticity of the Intermediate NBs to Flexible NBs is 14.37 which is in the range of the ratio of the measured GP of the Intermediate to Flexible shells of 11.07$\pm$8.85, and ratio of the elasticity of the Stiff NBs to Intermediate NBs is 1.37 which correlates well with the corresponding ratio of GPs which is  1.32$\pm$0.54.\\ 
The effect of Gly and PG on the properties of PL membranes for biological and biomedical applications has been extensively studied through experiments, numerical simulations, and molecular dynamics simulations \cite{33,35,40,41,42,43,65}. Gly is a good osmotropic agent enhancing the water-water H-bonding at the PL solvation shell and thereby imparting an ordering effect on PL packing \cite{41,62,66}. PG, on the other hand, is a synthetic molecule with lower polarity as compared to Gly \cite{67,68}. The lower polarity of PG also implies that it can be incorporated in the PL membrane through solvation of the head group, partitioning of PG into the hydrophobic core, or combination thereof as shown by Harvey et al. \cite{39}.  Furthermore, incorporation of PG results in a decrease in gel-liquid phase transition temperature of the acyl chains \cite{39}. The decrease in stiffness of PL membrane upon incorporation of PG has been utilized in formulations for ultra-deformable liposomes as an edge activator \cite{69}. Here, we show for the first time that addition of Gly and PG changes the shell structure and therefore acoustic behavior of the narrow size disperse NBs. The changes in shell in the shell properties are quantified both experimentally and numerically.\\
\textbf{2-Increased shell rupture threshold}\\
Gly stiffens the NB membrane (Fig. 4c), which limits the NB oscillation amplitude. Moreover, the stiffer shells need higher pressures for rupture \cite{18}. As soon as the shell ruptures, the amplitude of bubble oscillations increases significantly resulting in the enhancement of the NB scattered pressure \cite{18}. The shell resists the rupture until the applied pressure reaches a threshold at which tensile stresses on the shell exceeds the rupture threshold \cite{18}. The stiffer the bubble, the higher the rupture surface tension and consequently higher pressures are required to achieve the enhancement. The stiffening effect of Gly on the PL membrane has been well-established in literature. Recently, Abou-Saleh et al., reported that Gly induces water structuring around the PL membrane of a MB through the formation of a glassy layer that increases MB stiffness. The stiffening effect of Gly on the MB membrane was determined through the compression of a MB using a tip-less atomic force microscopy cantilever. The force to achieve a given compression was shown to increase with increasing Gly content up to 20$\%$ Gly \cite{65}. Conversely, PG softens the membrane making it more flexible and thereby requiring a lower $P_t$. PG has been used as an edge-activator for ultradeformable liposomes for enhanced drug delivery, especially through the skin \cite{70,71,72}. Ultradeformable liposomes have been shown to squeeze through narrow openings without disruption of its vesicular structure and this is facilitated by its flexible and strain-compliant membrane. Zhao et al. utilized drug-loaded liposomes with PG for enhanced delivery of epirubicin into breast cancer tumors \cite{37}. PG was specifically chosen for this study because PG-liposomes have higher encapsulation efficiency, better membrane flexibility, and longer stability as compared to normal liposomes.\\
Numerical simulations using the Marmottant model confirm that higher pressures are required for NB scattering enhancement when the NBs have stiffer shells \cite{18,57}. The shell can withstand finite tensions only; increasing the acoustic pressure gradually shows a strong abrupt enhancement above a critical pressure. This is due to the shell rupture: in this new state, the bubble oscillates as a free bubble. This is because above a critical tension (corresponding to $\sigma_{rupture}$), the shell ruptures and that part of the bubble surface is uncovered \cite{18}. Once this threshold has been reached, the surface tension upper bound will be the surface tension of water, allowing the bubble to expand more easily (which translates in the backscatter enhancement). The stiffer bubbles have more resistant shells, thus the rupture occurs at higher pressures. Numerical simulations predicted $\sigma_{rupture}$ of 0.0725 N/m, 0.084 N/m and 0.087 N/m for the flexible, intermediate and stiff shell NBs.\\
\textbf{3-Decreased initial surface tension}\\
The second reason behind the increase in the pressure threshold of the enhancement of the signal with the addition of Gly is the increased stability of the NBs with initial surface tension reduction. We have previously shown that the initial surface tension of the NBs decreases significantly ($p < 0.0001$) through the incorporation of Pluronic L10 \cite{73}. The initial surface tension decreased by 28\%  at a  lipid to Pluronic ratio of 0.2 \cite{73}. Here, addition of Gly has a similar stabilizing effect to incorporation of Pluronic by reducing the initial surface tension. Predictions of the numerical simulations validate this hypothesis as the predicted initial surface tension decreased from 0.067 N/m for the flexible NBs to 0.055 N/m for the stiff NBs. According to the numerical simulations, the differences between the initial surface tension and the surface tension for rupture determines the pressure threshold for the sudden signal amplification; the pressure threshold increases with increasing the margin between the initial and rupture surface tension.\\
\textbf{4-Increased shell viscosity}\\
Incorporation of Gly increased the viscosity of the shell from 0.9 nkg/s  for flexible NBs to 9 nkg/s for intermediate and 25 nkg/s for stiff NBs.  This can be explained by the higher viscosity of the Gly (1.412  Pa.s \cite{74}) compared to PG (0.042 Pa.s \cite{75}). Viscosity of a mixture of liquids can be calculated using \cite{C}:
\begin{equation}
\mu=x_a*\mu_a^{1/3}+x_b*\mu_b^{1/3}
\end{equation}
where x is the mass fraction,  $\mu$ is the viscosity and sub index a and b represent fluid a and b respectively. By neglecting the influence of lipids due to their small mass fraction and assuming the viscosity of 0.001 for PBS and densities of 1 g/ml for PBS, 1.04 g/ml for PG and 1.26g/ml for glycerol we can estimate the viscosity of each mixture as: $\mu_{flexible}$=0.0032 Pa.s, $\mu_{intermediate}$=0.0158 Pa.s and  $\mu_{stiff}$=0.0415 Pa.s. Thus $\frac{\mu_{intermediate}}{\mu_{flexible}}=4.94$ and $\frac{\mu_{stiff}}{\mu_{intermediate}}=2.63$ which correlates well with the ratio of the numerically fitted shell viscosities of $\frac{k_s^{intermediate}}{k_s^{flexible}}=5.3$ and $\frac{k_s^{intermediate}}{k_s^{flexible}}=2.77$.\\
Despite the good quantitative agreement between experiments and simulations, however, the goal of the simulations in this paper is to shed insight on the physical mechanisms of the NB behavior with different shells and the threshold behavior observed in the experiments. The simulation parameters that are presented as best fit to each case are are representative of the relative comparison between shell parameters but the absolute value for each parameter may not be accurate. Accurate quantification of the physical parameters of the NBs is a challenging task and requires attenuation and scattering measurements in tandem. Nevertheless, the estimated values for the NB shell parameters here, are consistent with the reported values for MBs with similar shell compositions \cite{D,E,F} (using linear estimations) and parameters that were extracted using optical measurements of radius-time curves \cite{57} and pressure dependent attenuation measurements\cite{G,H}. \\
The use of NBs with a narrow size distribution in this study significantly aided in observing the effect of the shell structure on the bubble behavior \cite{51}. Such a clear difference in the behavior of various shelled bubbles has not been observed to date likely due to the absence of size-controlled measurements. The polydispersity of MBs may be the reason behind why there was no clear difference between the acoustic behaviors of different shell MBs in \cite{51} with different GP values.  This shows the importance of the applications of monodisprse NBs and MBs to achieve high control over their acoustic behavior, making the therapeutic and imaging effects more potent while at the same time increasing the safety of medical procedures. These findings, further confirm the results of previous studies on the importance of narrow size distributions of UCAs on their response to ultrasonic exposure \cite{76}. Here, we show for the first time that the acoustic response of narrow sized NBs can be controlled and altered by their shell structure. 
The controllable pressure threshold in this study has potential advantages for ultrasound contrast enhanced methods that rely on the nonlinear response of UCAs. One of these techniques is amplitude modulation where two pulses with different pressure amplitude are used in the imaging sequence. One pulse usually has an amplitude that is twice the other pulse. The received signals are scaled and subtracted upon receive. Due to the linear response of the tissue, the signal from tissue cancels and the only remaining signal is from UCAs, increasing the contrast to tissue (CTR). Sending a pulse below the pressure threshold and sending one above the threshold for enhancement will significantly increase the CTR. An increase in CTR would be particularly beneficial in ultrasound molecular imaging. Sojahrood and Kolios numerically investigated the pressure dependent super-harmonic resonances of monodisperse UCAs and showed that, above a pressure threshold, a significant increase in harmonic emissions is expected \cite{76}. This can aid in heating enhancement in treatments while reducing the undesired effects in the off-target tissue. The dynamics of size isolated UCAs which are excited by their pressure dependent resonance frequency (PDfr) has also been numerically investigated \cite{61}. Above a pressure threshold, bubble oscillations undergo an abrupt increase, resulting in the enhancement of the non-destructive scattered pressure by the bubbles. The authors concluded that the use of PDfr can used to increase the contrast in amplitude modulation imaging-based techniques. Moreover, the attenuation of the UCAs in the beam path can be suppressed to allow more ultrasound energy to reach bubbles at the target. Therefore, eliminating the effects of size disparity in bubble populations is a highly effective method, in principle, to enhance and control the outcome of the diagnostic and therapeutic procedures. In agreement with conclusions of \cite{61,77}, the reduction of pre-focal beam attenuation has been experimentally shown in \cite{76} where monodisperse populations of lipid coated MBs were sonicated by their PDfr.
We showed that NBs with flexible shells need smaller amplitude acoustic pressures for the non-linear oscillations leading to the pressure dependent scattering enhancement. This leads to a higher scattering cross section and thus better outcomes for imaging. Stiffer shells increase the pressure to higher values thus making them more suitable for therapeutic purposes like enhanced heating applications where higher pressures are required \cite{78}. Importantly, due to the negligible oscillation amplitude of the pre-focal NBs, and taking advantage of the steep pressure gradients of focused ultrasound transducers, we may significantly decrease the attenuation of pre-focal NBs in the ultrasound path. Thus, delivering energy to the resonant NBs at the target will contribute to efficiently producing enhanced heating effects. Undesired heating in the off-target region is minimized due to the off resonant bubbles. 
\section{Conclusion}
NBs of narrow size distribution with three different shell compositions were manufactured. The relative shell stiffness of different NB formulations was assessed by calculating the average GP value from the relative fluorescence intensities at 450-nm and 500-nm using a two-photon excitation microscopy technique. NBs prepared with 20 wt$\%$ Gly show the highest GP and therefore have the highest shell stiffness, while NBs prepared with 20 wt$\%$ of PG show the lowest GP and therefore have the lowest shell stiffness. We introduced a simple and efficient method by which high concentrations of narrow-sized NBs can be prepared through filtration for its use in ultrasound imaging experiments. Acoustic measurements of signals from filtered NBs showed that the difference in shell stiffness has a pronounced effect in the pressure threshold $P_t$ of PL-stabilized NB solution, with the flexible membrane requiring lower PNP and stiffer membrane requiring higher PNP to elicit nonlinear oscillations. Numerical simulations confirmed the experimental observations of the stiffness dependent threshold behavior. 
\section{Acknowledgement}
Research reported in this publication was supported by the National Institute of Biomedical Imaging and Bioengineering of the National Institutes of Health under award number R01EB025741 and the Office of the Assistant Secretary of Defense for Health Affairs, through the Prostate Cancer. Research Program under Award No. W81XWH-16-1-0371. Views and opinions of, and endorsements by the author(s) do not reflect those of the National Institutes of Health or of the Department of Defense. AJ Sojahrood was supported by CIHR vanier scholarship. M.C.K. and A.S. received support from the CIHR and NSERC. P.W. and E. P. thank NSF CAREER award $\#$1551943 for financial support. Al de Leon and Agata Exner would like to acknowledge the help from Olive Jung.

\end{document}